\documentclass[aps,prd,amsmath,groupedaddress]{revtex4}
\usepackage{epsfig}
\usepackage{graphicx}

\begin{document}

\title{Effect of tetrahedral shapes in heavy and superheavy nuclei}

\author{P.~Jachimowicz}
 \affiliation{Institute of Physics,
University of Zielona G\'{o}ra, Szafrana 4a, 65-516 Zielona
G\'{o}ra, Poland}

\author{M.~Kowal} \email{michal.kowal@ncbj.gov.pl}
\affiliation{National Centre for Nuclear Research, Ho\.za 69,
00-681 Warsaw, Poland}

\author{J.~Skalski}
\affiliation{National Centre for Nuclear Research, Ho\.za 69,
00-681 Warsaw, Poland}

\date{\today}

\begin{abstract}
{\noindent
 We search for effects of tetrahedral deformation $\beta_{32}$
 over a range of $\sim 3000$ heavy and superheavy nuclei, $82\leq Z \leq 126$,
 using a microscopic-macroscopic model based on the deformed Woods-Saxon
 potential, well tested in the region.
 We look for the energy minima with a non-zero tetrahedral distortion, both
 absolute and conditional - with the quadrupole distortion constrained to
 zero. In order to assure reliability of our results
 we include 10 most important deformation parameters in the energy
 minimization.  We could not find any cases of stable tetrahedral shapes.
 The only sizable - up to 0.7 MeV - lowering of the ground state
 occurs in superheavy nuclei $Z\geq 120$ for $N=173-188$, as a result of
 a {\it combined} action of two octupole deformations: $\beta_{32}$ and
 $\beta_{30}$, in the ratio $\beta_{32}/\beta_{30}\approx \sqrt{3/5}$.
 The resulting shapes are moderately oblate, with the superimposed distortion
 $\beta_{33}$ {\it with respect to the oblate axis},
 which makes the equator of the oblate spheroid slightly triangular.
 Almost all found conditional minima are excited and not protected by any
 barrier; a handful of them are degenerate with the axial minima.
 }
\end{abstract}
\maketitle

\section{Introduction}

 The idea of intrinsic shape of a nucleus turned out instrumental for
 understanding many features of the nuclear structure and spectroscopy.
 In particular, specific nuclear shapes were related to the prominent shell
  effects in both proton and neutron systems exhibited by the nuclear
  binding, and to the observed patterns of collective excitations.
  Besides the axial quadrupole distortion which is the nuclear
 deformation of primary importance, the secondary effects of hexadecapole
 \cite{B40exp1,B40exp2} and, in some regions, octupole
 \cite{B30exp1,B30exp2,B30theory} distortion are clearly recognized.
 Additionally, there are theoretical predictions of quadrupole triaxial
  equilibrium shapes in some nuclei, e.g. \cite{B22theory1,B22theory2}, but
   rather limited experimental evidence for them, see e.g.
 \cite{Ober2009,B22exp}.
 From the theoretical point of view even more exotic shapes are possible,
 characterized by a high rank symmetry group which would lead to an extra
 degeneracy of s.p. energy levels. One such possibility is the tetrahedral
 symmetry. It is well known that many quantum
 objects, like molecules, fullerenes and alkali metal clusters prefer such a
 shape in their ground state.
 Due to these facts a hypothesis of tetrahedral symmetry of an atomic nucleus
 was put forward as early as in the 1970s for $^{16}$O
 \cite{B32_light1,B32_light2} in relation to its expected four-$\alpha$ cluster
  structure.
 Since the 1990s, such concept has been extended also to the heavier systems,
 e.g. \cite{first1,first2} and then intensively studied, both within
 microscopic-macroscopic (MM) \cite{MMandSF1,MMandSF2,MM1,MM2,JRKSS} and
 selfconsistent models \cite{MMandSF1,MMandSF2,SF0,SF1,SF2,SF3,SF4,SF5}.
 Generally,
 these studies are inconclusive since: a) the existence of global tetrahedral
 minima was rare and model-dependent b) contradictory results were obtained
 within the same models. Similar ambiguity occurs also in experiments which
 so far either did not give a clear evidence \cite{expB32_1} or even gave a
 strong evidence against tetrahedral symmetry \cite{expB32_2,expB32_3}.
 For example, negative-parity bands in $^{156}$Dy, observed quite recently
 \cite{expB32_4}, are most likely related to the octupole excitations
 rather than the exotic tetrahedral symmetry.

 Here we summarize the results of a search for tetrahedral minima in heavy and
 super-heavy nuclei obtained within the MM model based on the deformed
 Woods-Saxon potential with parameters used many times before, therefore well
 tested in this region.
 The present work is a much improved version of \cite{JRKSS}, extended to
 odd-$A$ and odd-odd nuclei, with an expanded space of deformations used for
 searching ground-state (absolute) tetrahedral minima.

 \section{Calculations}

 The microscopic-macroscopic results were obtained with the
   deformed Woods-Saxon potential. The nuclear deformation enters via a
  definition of the nuclear surface \cite{WS}:
  \begin{eqnarray}
  \label{shape}
    R(\theta,\varphi)&=& c(\{\beta\}) R_0 \{ 1+\sum_{\lambda>1}\beta_{\lambda 0}
   Y_{\lambda 0}(\theta,\varphi)+
\nonumber \\
&&
    \sum_{\lambda>1, \mu>0, even}
   \beta_{\lambda \mu} Y^c_{\lambda \mu} (\theta,\varphi)\}  ,
  \end{eqnarray}
  where $c(\{\beta\})$ is the volume-fixing factor. The real-valued spherical
  harmonics $Y^c_{\lambda \mu}$, with even $\mu>0$, are defined in terms of the
  usual ones as: $Y^c_{\lambda \mu}=(Y_{\lambda \mu}+Y_{\lambda -\mu})/ \sqrt{2}$.
  In other words, we consider shapes with two symmetry planes. Note, that
 traditional quadrupole deformations $\beta$ and $\gamma$ are related to
 $\beta_{20}$ and $\beta_{22}$ by:
   $\beta_{20}=\beta\cos\gamma$ and $\beta_{22}=\beta\sin\gamma$.

 The $n_{p}=450$ lowest proton levels and $n_{n}=550$ lowest
 neutron levels from the $N_{max}=19$ lowest shells of the deformed oscillator
 were taken into account in the diagonalization procedure.
 For the macroscopic part we used the Yukawa plus
 exponential model \cite{KN}.

 All parameters used in the present work,
 determining the s.p. potential, the pairing strength, and the
 macroscopic energy, are equal to those used previously in the calculations
 of masses \cite{WSpar} and fission barriers \cite{Kow,2bar,bar2017} in
 actinides and the heaviest nuclei. In particular, we took the
 "universal set" of potential parameters and the pairing strengths
 $G_n=(17.67-13.11\cdot I)/A$ for neutrons, $G_p=(13.40+44.89\cdot I)/A$
 for protons ($I=(N-Z)/A$).
 As always within this model, $N$ neutron and
 $Z$ proton s.p. levels have been included when solving BCS equations.
 For systems with odd proton or neutron (or both) we use blocking. We assume
 the g.s. configuration consisting of an odd particle
 occupying one of the levels close to the Fermi level and the rest of the
 particles forming a paired BCS state on the remaining levels.
 Any minimum, including the ground state, is found by minimizing over
 configurations, blocking particles on levels from the 10-th below to 10-th
 above the Fermi level.

 We performed three types of calculations looking for both conditional
 and ground-state tetrahedral minima in nearly 3000 heavy and superheavy
 nuclei with $Z\geq 82$.

 1) Conditional tetrahedral minima were found by
 fixing quadrupole deformations at zero: $\beta_{20}=\beta_{22}=0$, and
 calculating total energy with the step 0.02 in $\beta_{32}$
 by minimization over the other seven deformation parameters:
 $\beta_{30}$, $\beta_{40}$, $\beta_{42}$, $\beta_{50}$, $\beta_{60}$, $\beta_{70}$, $\beta_{80}$.
 The occurence of a  minimum at $\beta_{32}\neq 0$ in such an energy plot
 (after additional interpolation of the energy to the step 0.01 in $\beta_{32}$) signals the conditional
 minimum, usually excited above the g.s.
 The rationale behind this procedure is that, as known from other studies,
  quadrupole deformation does not cooperate with the tetrahedral one \cite{MM1,MMandSF2},
  and switching off the effects of the quadrupole might help to locate a prominent
 tetrahedral shell effect at sizable deformation $\beta_{32}$.

 2) The ground states in all nuclei were found initially by the minimization
 over seven axial deformations $\beta_{\lambda 0}$, $\lambda=2$ - 8.

 3) Finally, the ground states were found for the second time, by the
 minimization over ten deformations: the axial ones from 2) plus $\beta_{22}$,
 $\beta_{32}$, and $\beta_{42}$.

  In additional calculations, for a restricted region of SH nuclei in which
  the minima found in 3) were $\sim 0.5$ MeV deeper than those resulting
 from 2), we minimized energy with respect to nine deformations, excluding
  $\beta_{32}$. The aim was to see whether the effect is driven by
  $\beta_{32}$.

  In all calculations we used one non-axial version of the WS code to
 eliminate possible numerical differences which could follow from the different
  imposed spatial symmetries.
 When looking for ground state minima, a minimization for each nucleus
 was repeated at least 30 times, with various starting points,
 in order to ensure that the proper ground state was found.
 As, especially for superheavy nuclei,
 the minimization can end behind the fission barrier, only minima within the
 the fission barrier were accepted.

 \section{Results and Discussion}

 \subsection{Tetrahedral minima}

 The map of tetrahedral deformations in the obtained conditional
 minima is shown in Fig. 1. We emphasize that in these minima the quadrupole
 deformation was forced to vanish in order to exhibit large tetrahedral shell
 effects. As may be seen, the largest $\beta_{32}$ reach $\sim0.2$.
 The conditional tetrahedral minima with sizable $\beta_{32}>0.1$ occur
 in three regions: a wide region around $Z=94$, $N=136$, and two very exotic
 regions: $Z\approx 98$, $N\approx 192$, and $Z=126$, $N\approx 192$.
 This, however, should be confronted with the excitation energies
 of the conditional minima above the axially symmetric g.s. minima [found in
 calculations no. 2)], shown in Fig. 2. As may be seen there, the low excitation
  energies in the {\it a priori} interesting first region occur in neutron rich
 $Z=84-94$ nuclei and in the very neutron deficient $Z=92-106$ isotopes
 which probably cannot be reached in experiment.
  There are some very exotic nuclei in which the conditional tetrahedral
 minima lie lower than the axially symmetric ones, but the largest difference
   is only 0.25 MeV. The conclusion from this part of the study is that in the
 whole investigated region there are no prominent low-energy tetrahedral minima.

\begin{figure}
\begin{minipage}[t]{120mm}
\centerline{\includegraphics[scale=0.5]{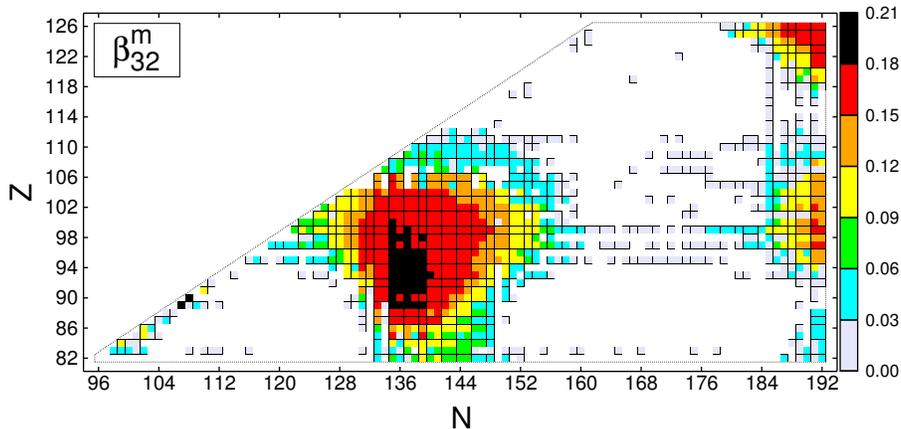}}
\end{minipage}
\caption{{\protect (color online) Deformation $\beta_{32}$ at the conditional minima obtained
by setting $\beta_{20}=\beta_{22}=0$; the minimization performed over 7 other deformation parameters (see text).}}
\label{fig1}
\end{figure}

\begin{figure}
\begin{minipage}[t]{120mm}
\centerline{\includegraphics[scale=0.5]{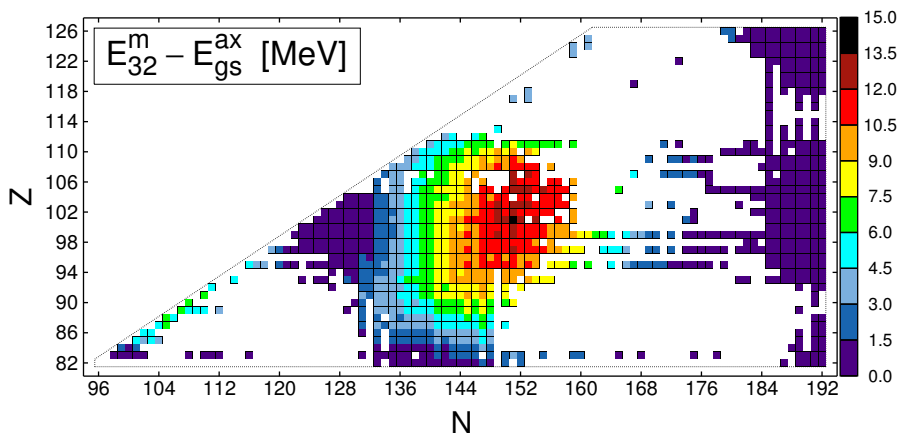}}
\end{minipage}
\caption{{\protect (color online) Excitation energy of the conditional minima with $\beta_{32}\neq 0$ above the
axially symmetric g.s. }}
\label{fig2}
\end{figure}

  Among the group of neutron-rich $Z=84$ - 94 nuclei, there are altogether
 fourteen conditional minima below 2 MeV excitation energy: nine in Po isotopes,
 four in Rn isotopes, and in $^{223}$Np.
 An example of the energy landscape of a nucleus
 $^{219}$Po, with coexisting shallow, nearly degenerate minima: with
 $\beta_{32}\approx 0.1$ and the wide prolate-$\beta_{30}$ one, is shown in
 Fig. 3, in three
 projections: ($\beta_{20}$, $\beta_{22}$), ($\beta_{20}$, $\beta_{30}$), and
 ($\beta_{20}$, $\beta_{32}$). These maps were obtained by using all
 10 deformations by the minimization over 8 remaining ones. The unusual
 landscape may be interpreted as two competing minima with a slight barrier
 between them.

   Concerning the excited minima, the important question is whether they
   are protected by a barrier from the transition to the axially symmetric
  g.s. minimum.
   The typical situation is shown in Fig. 4, for the nucleus $^{222}$Rn.
  The landscape in ($\beta_{20}, \beta_{32}$) plane was obtained by the
 minimmization over 8 remaining deformations. One can see that the conditional
 minimum with $\beta_{20}=0$ is not a minimum after lifting the constraint on
 $\beta_{20}$: the very shallow real tetrahedral minimum occurs at
 $\beta_{20}=-0.04$ and $\beta_{32}=0.04$ which is smaller than
 $\beta_{32}=0.10$ of the conditional minimum. The barrier between the
 tetrahedral and the
  axial prolate g.s. minimum is less than 300 keV. One has to notice though,
 that the presented picture involves the minimization over other deformations,
 while finding the height of the saddle would require another method, like,
 for example, the imaginary water flow (e.g. \cite{MoBar,bar2017}), applied in the whole deformation
 hypercube.

\begin{figure}
\begin{minipage}[t]{180mm}
\hspace{-3mm}
\includegraphics[scale=0.3]{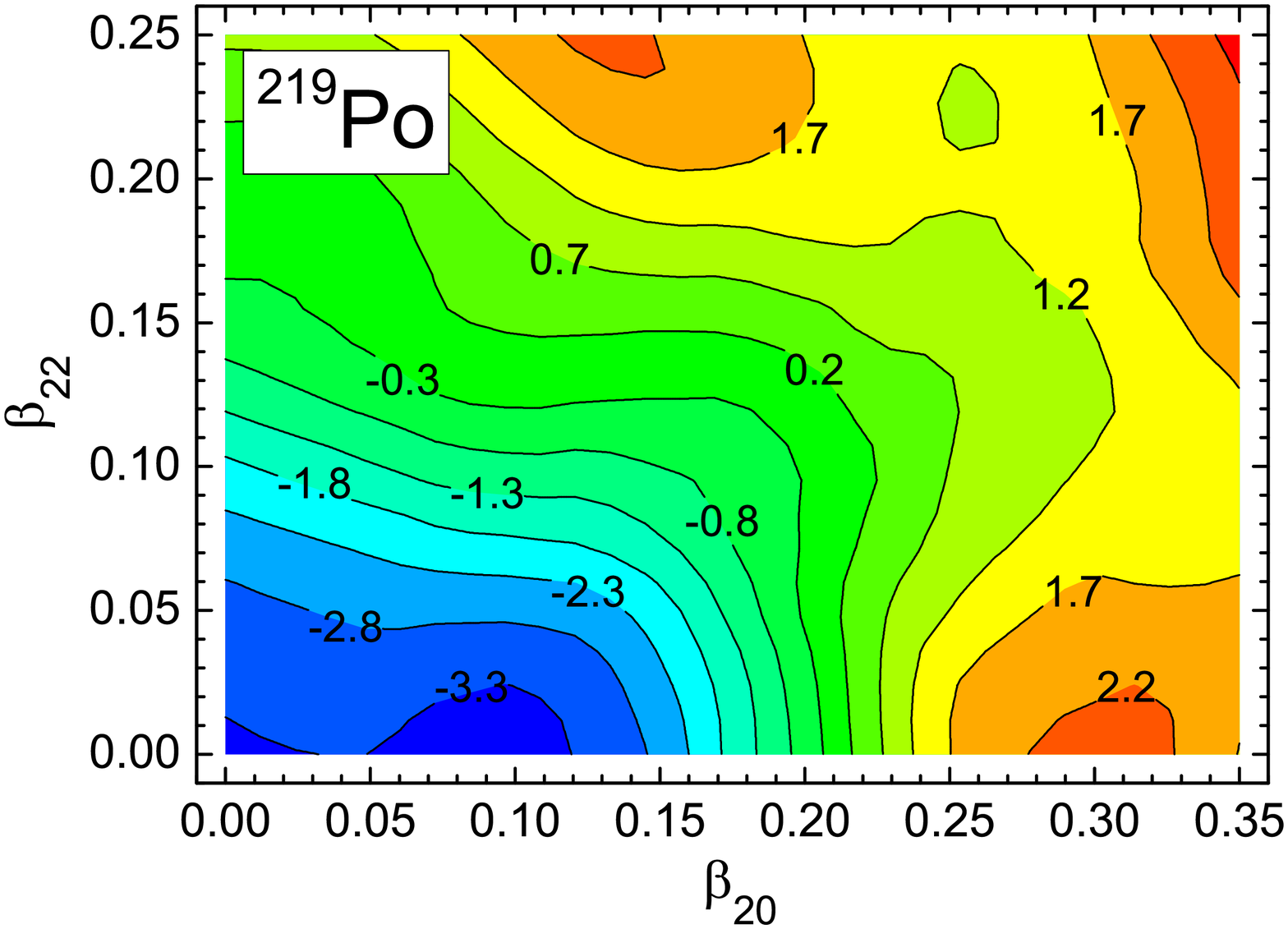}
\hspace{-6mm}
\includegraphics[scale=0.3]{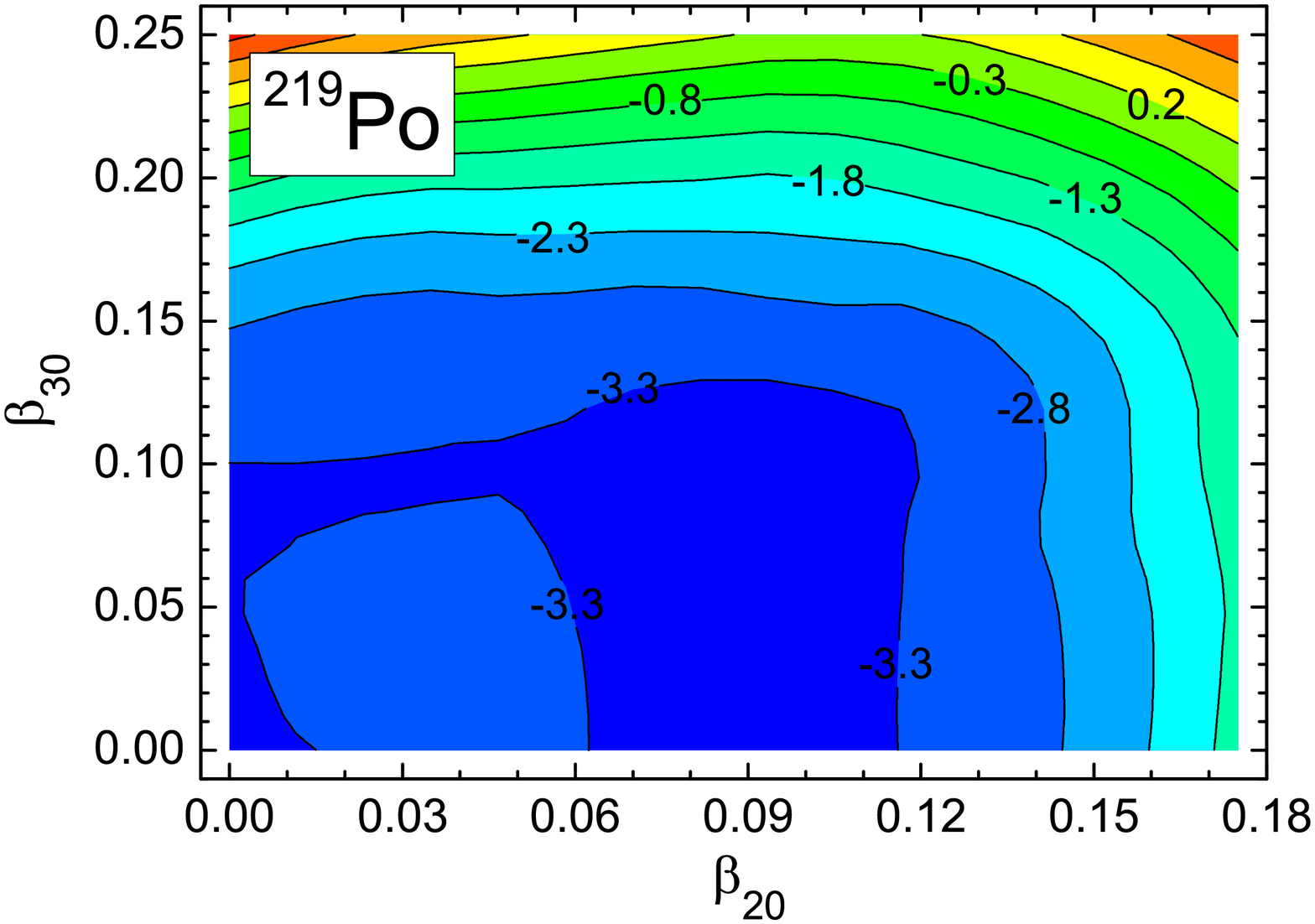}
\hspace{-3mm}
\includegraphics[scale=0.3]{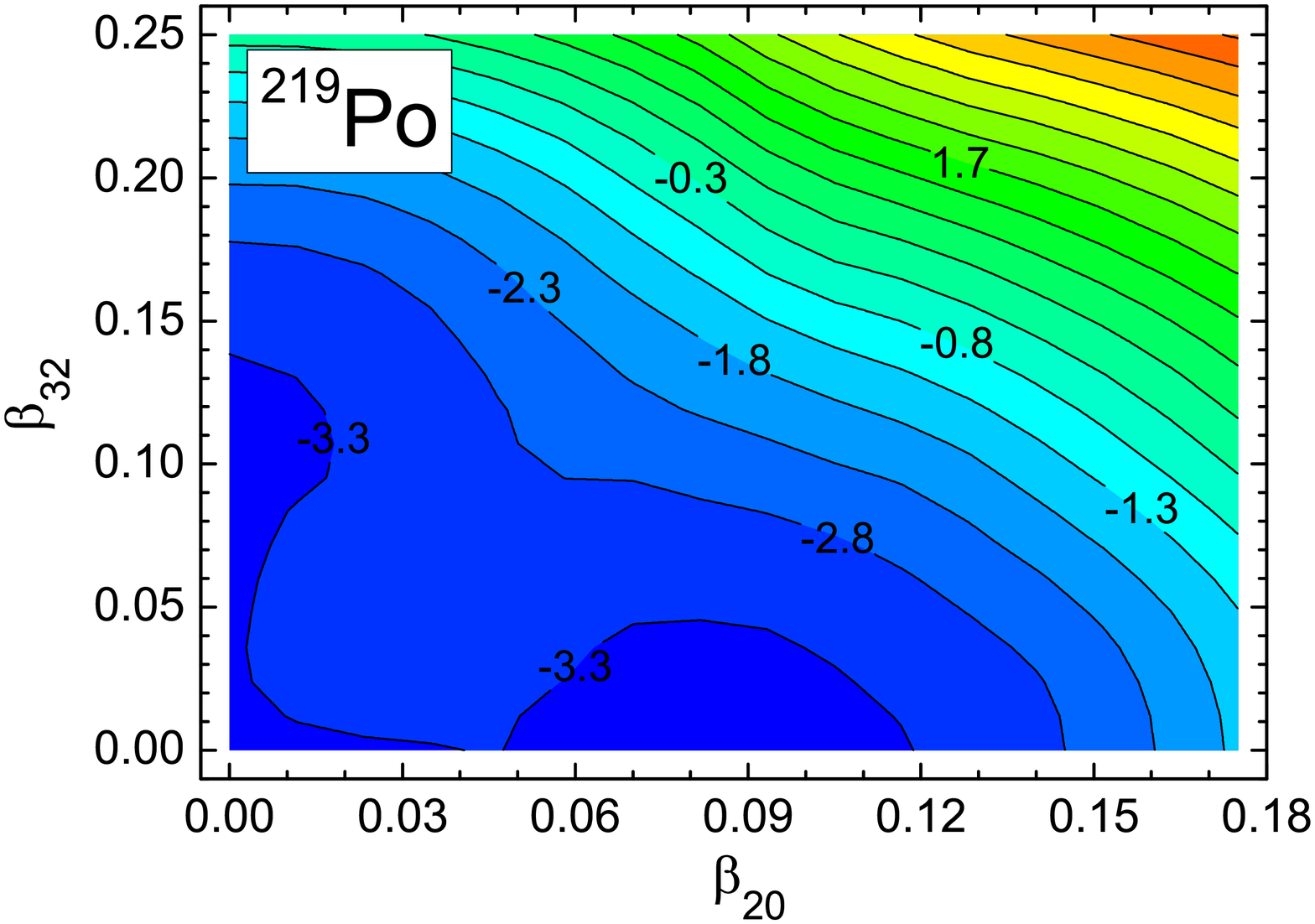}
\end{minipage}
\caption{{\protect (color online) Energy landscapes in $(\beta_{20}, \beta_{22})$, $(\beta_{20},
\beta_{30})$ and $(\beta_{20}, \beta_{32})$ planes for $^{219}$Po, calculated from the minimization over
the remaining eight deformation parameters. Energy calculated relative to the macroscopic energy at the spherical shape.}}
\label{figmap}
\end{figure}

\begin{figure}
\begin{minipage}[t]{120mm}
\centerline{\includegraphics[scale=0.3]{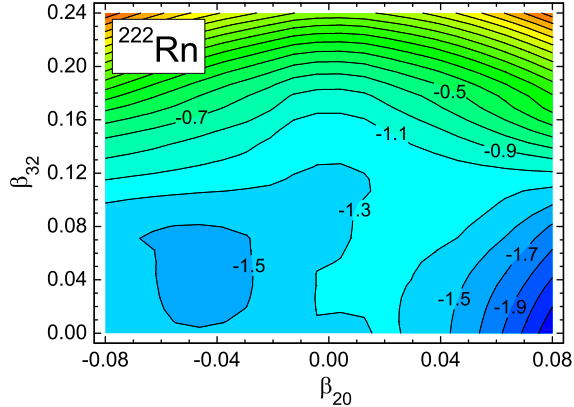}}

\end{minipage}
\caption{{\protect (color online) Energy map for $^{222}$Rn; only a small barrier may be seen between the
slightly oblate, $\beta_{32}=0.04$ minimum and the axially symmetric g.s. with $\beta_{20}\approx0.12$
(not fully visible on this map).}}
\label{fig7}
\end{figure}

 \subsection{Minima including tetrahedral deformation}

 In the next step we found all nuclei in which the energy minimization
 over 10 deformations, including nonaxial $\beta_{22}$, $\beta_{32}$ and
 $\beta_{42}$, lead to the g.s. lying lower than the axially symmetric
 minimum. They are shown in Fig. 5.
 In many of them, the effect comes entirely from the quadrupole and
 hexadecapole nonaxiality ($\beta_{22}$ and $\beta_{42}$). Such is the
 situation in nuclei with $Z<118$, forming vertical lines in Fig. 5:
  at $N=121$, 179 (nuclei with small oblate deformation $\beta_{20} > -0.1$),
 and $N=137$, 153 (well deformed prolate nuclei with $\beta_{20}\approx 0.25$).
 Among the last group, there are only a few cases in which a small deformation
 $\beta_{32}\approx 0.02 - 0.03$ occurs in the g.s. On the other hand,
 in many nearly spherical $N=185$ isotones a small distortion
 $\beta_{32}\approx 0.03$ results from the energy minimization.
 The energy differences greater than 200 keV between non-axial and axial
 minima occur in rather exotic nuclei.
 For example, the purely tetrahedral effect occurs in the neutron-poor
 Es isotope with $N=128$ neutrons and in ultra-neutron-rich
 Es isotopes with $N=185-192$, and also in a few nuclei around them.

\begin{figure}
\begin{minipage}[t]{120mm}
\centerline{\includegraphics[scale=0.5]{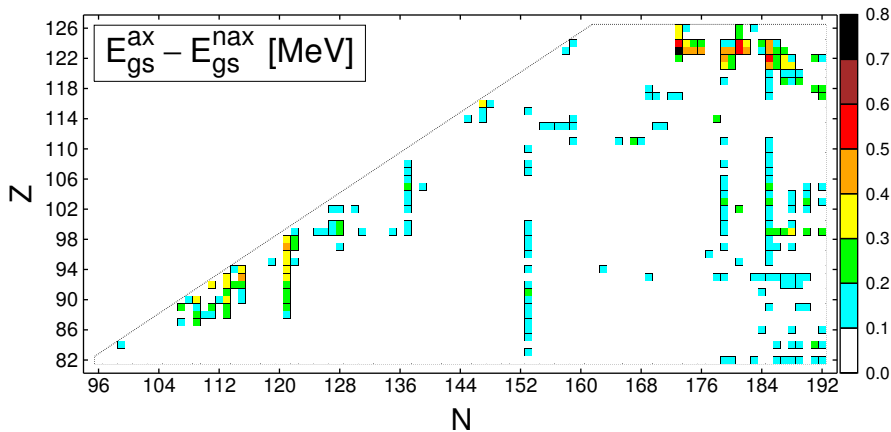}}
\end{minipage}
\caption{{\protect (color online) Nuclei for which the minima obtained by using 10 deformations
including quadrupole nonaxiality $\beta_{22}$ and octupole $\beta_{30}$ and $\beta_{32}$ are below the axial ones.}}
\label{fig6}
\end{figure}

\begin{figure}
\begin{minipage}[t]{120mm}
\centerline{\includegraphics[scale=0.5]{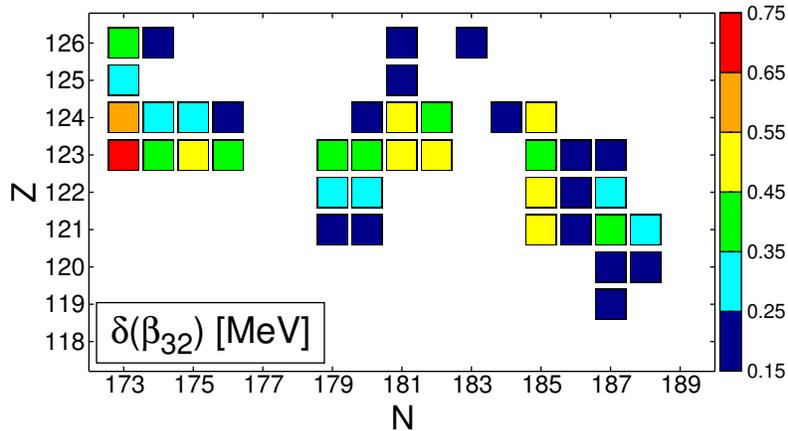}}
\end{minipage}
\caption{{\protect (color online) Nuclei around $^{296}123$ for which the inclusion of $\beta_{32}$ lowers
the g.s. by at least 150 keV as compared to the minimization over nine deformations.}}
\label{fig7}
\end{figure}

\begin{figure}
\begin{minipage}[t]{180mm}
\hspace{-3mm}
\includegraphics[scale=0.3]{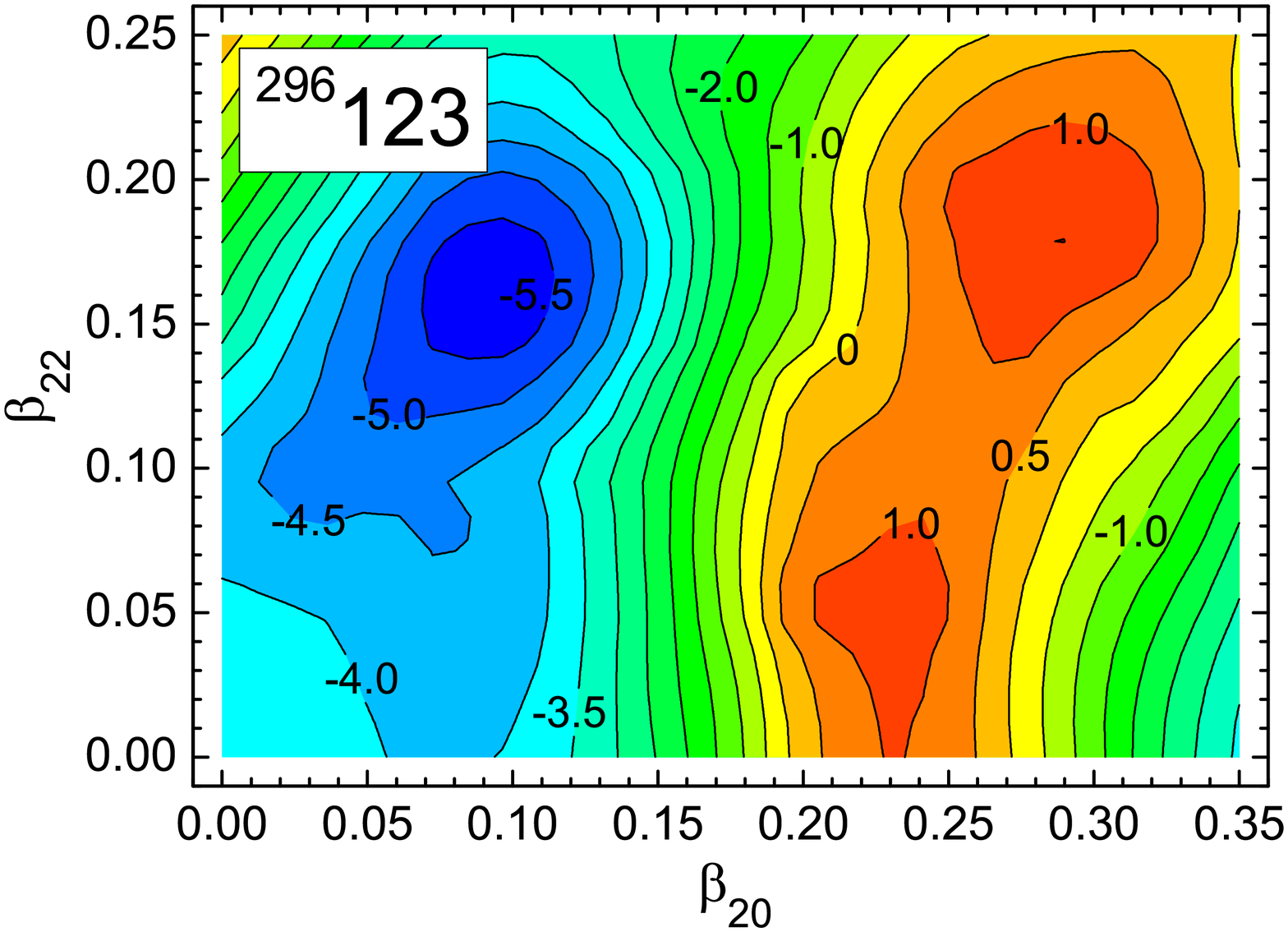}
\hspace{-6mm}
\includegraphics[scale=0.3]{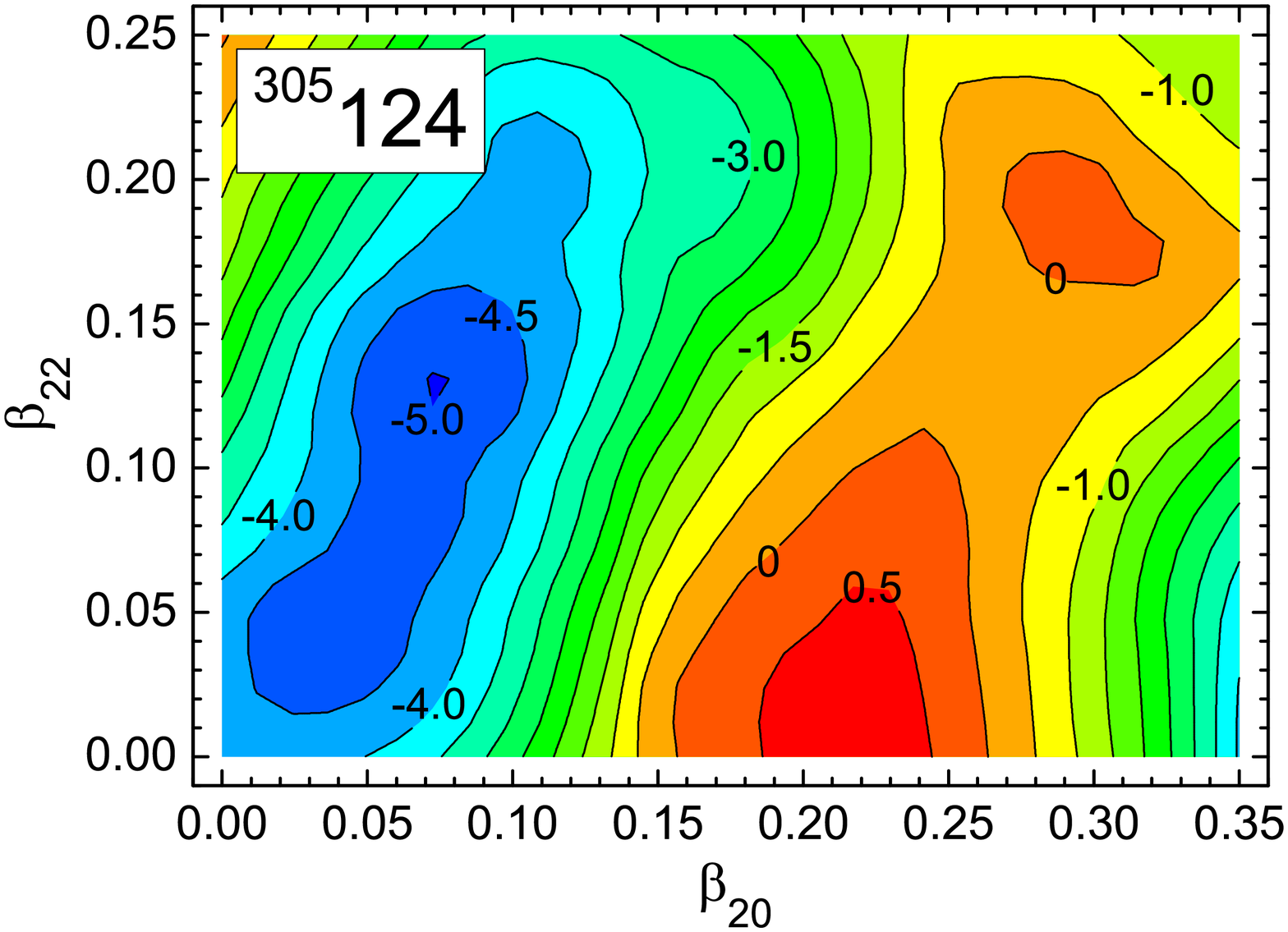}
\hspace{-3mm}
\includegraphics[scale=0.3]{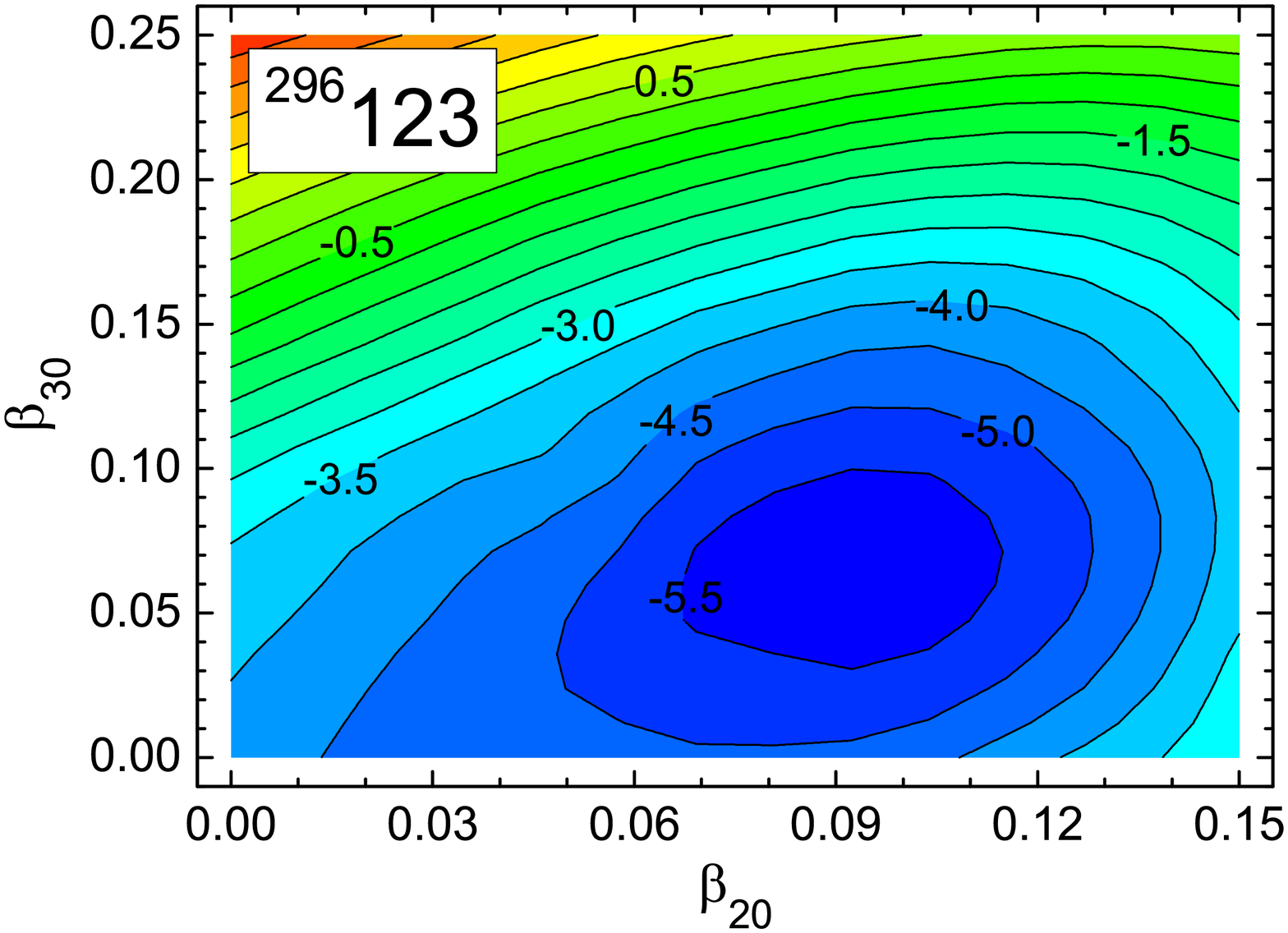}
\hspace{-6mm}
\includegraphics[scale=0.3]{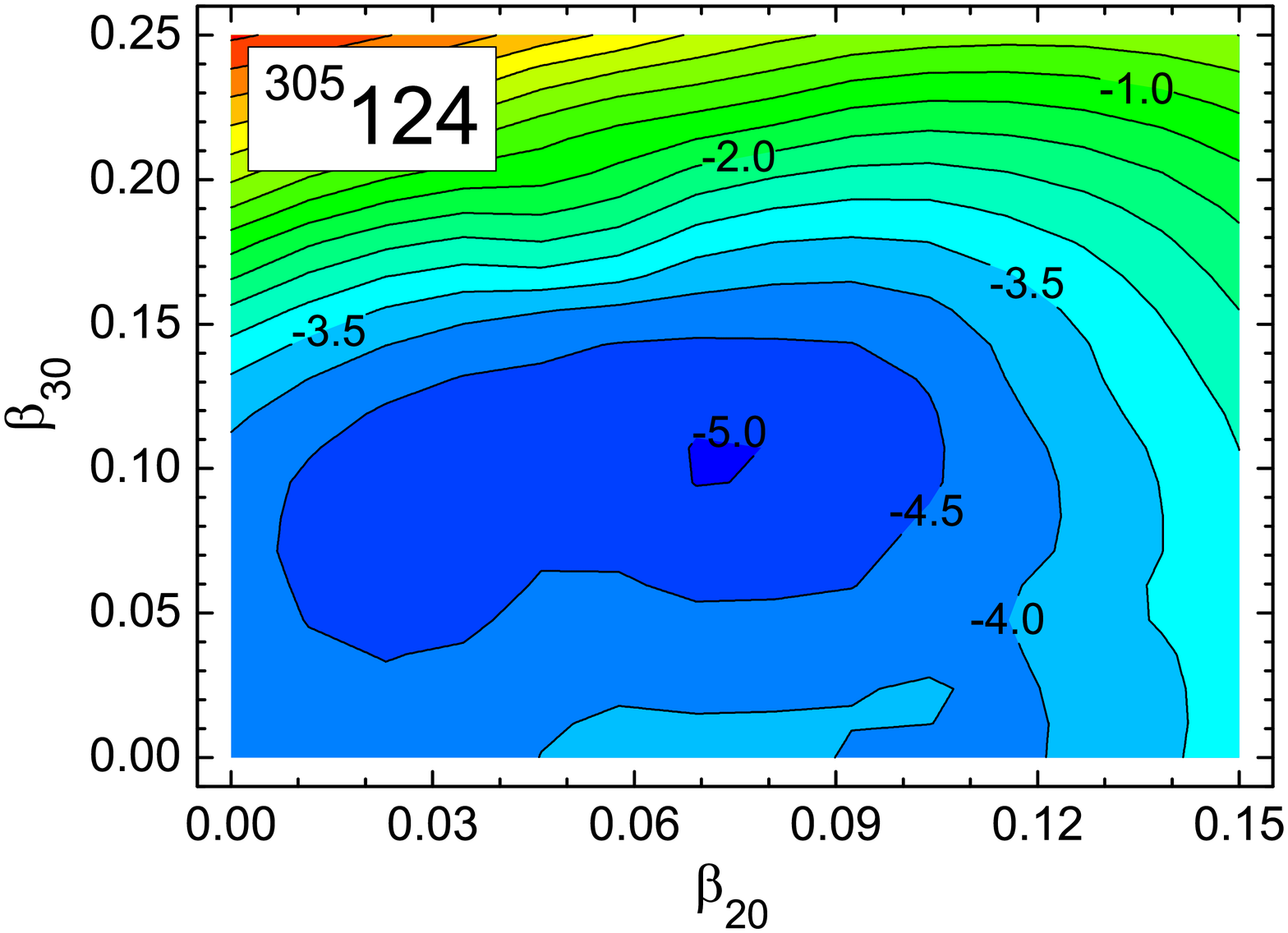}
\hspace{-3mm}
\includegraphics[scale=0.3]{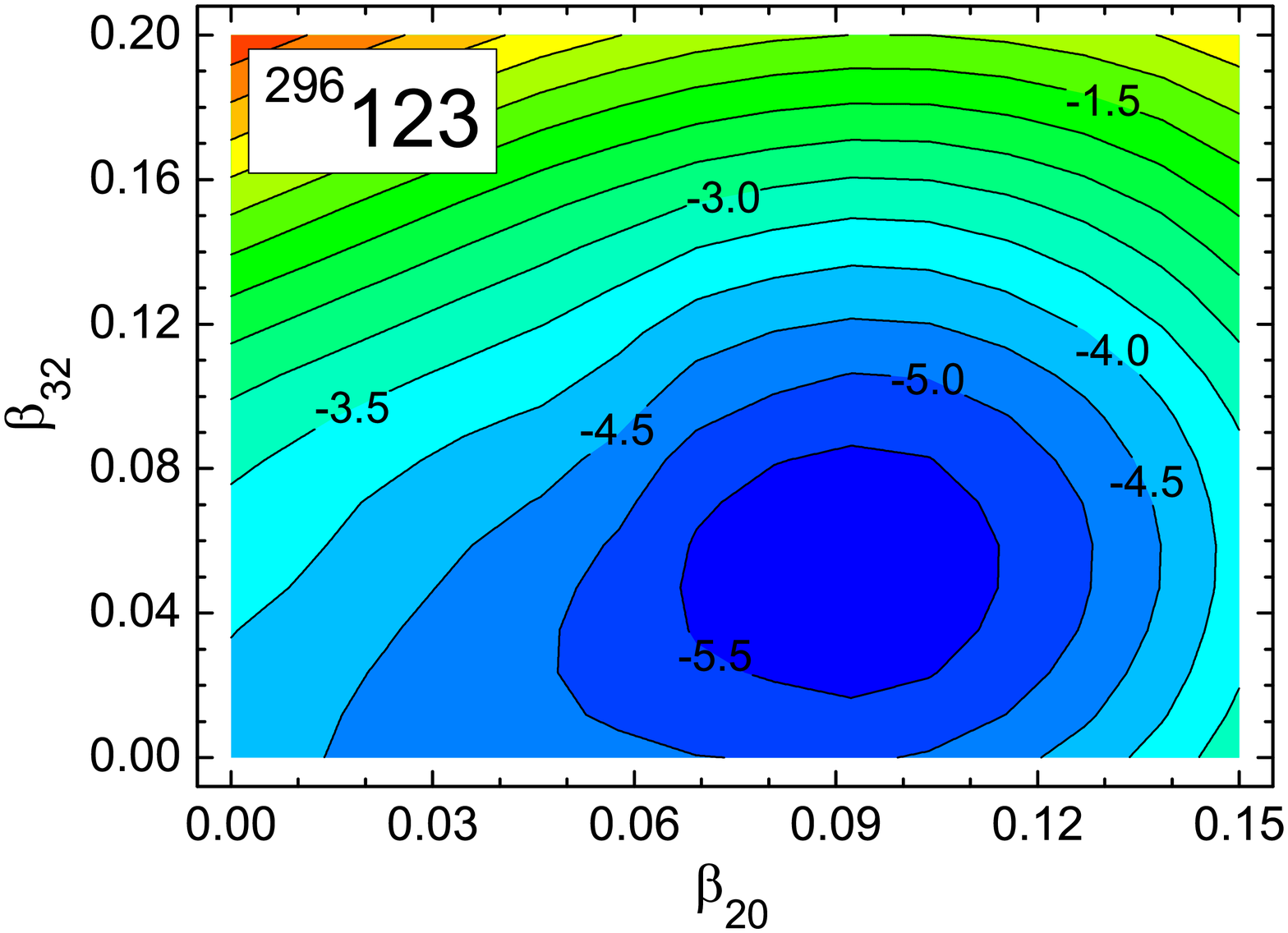}
\hspace{-6mm}
\includegraphics[scale=0.3]{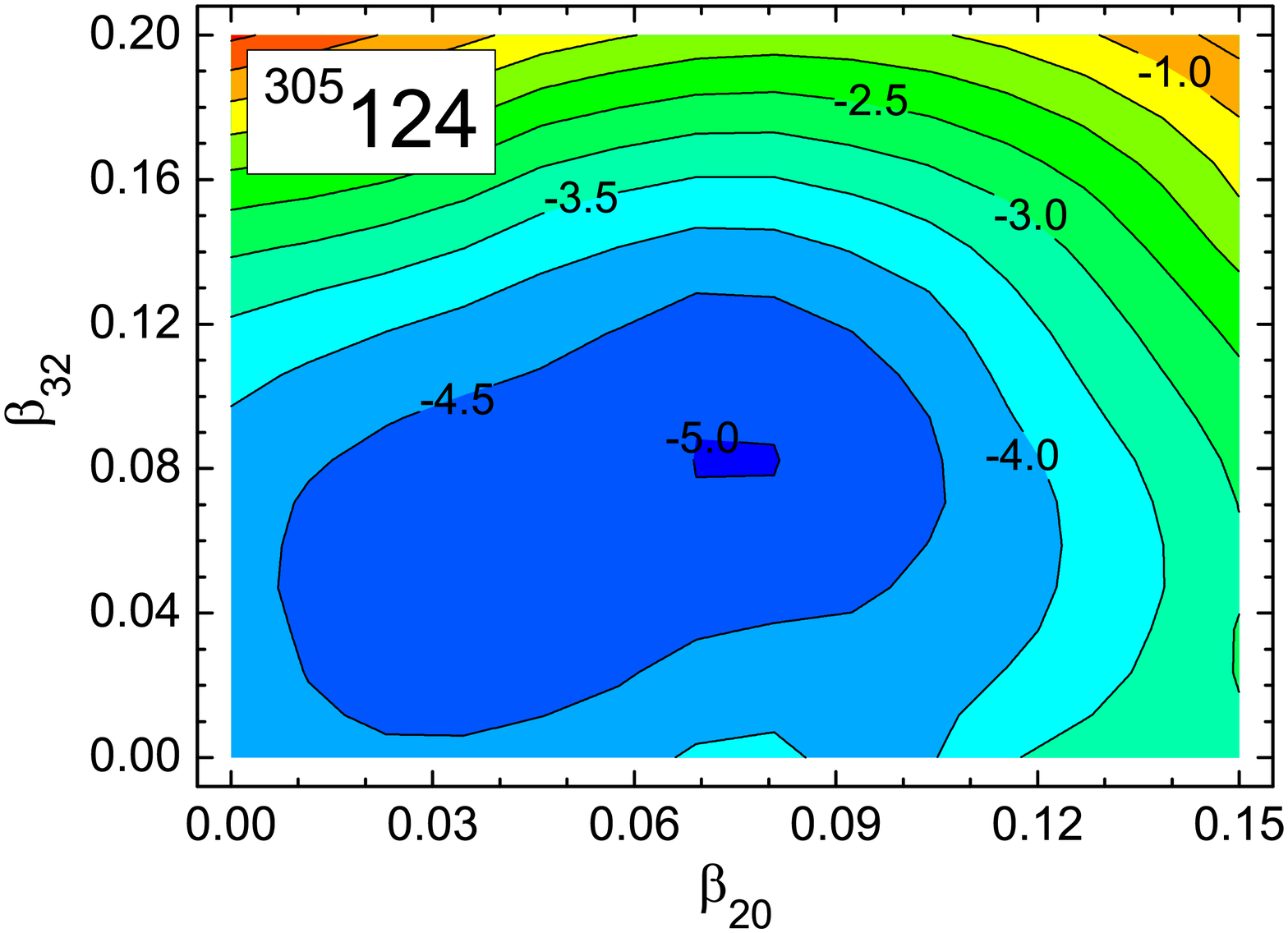}
\end{minipage}
\caption{{\protect (color online) Energy landscapes in $(\beta_{20},\beta_{22})$, $(\beta_{20}, \beta_{30})$
and $(\beta_{20}, \beta_{32})$ planes for nuclei: $Z=122$, $N=173$ and $Z=124$, $N=181$, calculated from the
minimization over the remaining eight deformations. Energy calculated relative to the macroscopic energy at
the spherical shape.}}
\label{figmap}
\end{figure}

 The largest effect occurs for SH nuclei with $Z>118$, especially around
 $Z=123$, $N=173$.
 In Fig. 6 are shown nuclei from this region in which the tetrahedral
 deformation $\beta_{32}$ lowers the ground state by more than 150 keV.
 This effect is calculated as the difference between energies in the g.s.
 minimum from the minimizations including nine (excluding
 $\beta_{32}$) and ten (including $\beta_{32}$) deformations. Although
 this could be named a "pure" $\beta_{32}$ effect, the reality is more
 intricate. It turns out that including terahedral deformation induces
  also oblate quadrupole and the {\it axial octupole} $\beta_{30}$.
 The obtained minima, corresponding to moderately oblate shapes with
  octupole distortions in the ratio: $\beta_{32}/\beta_{30}\approx \sqrt{3/5}$
 are equivalent to the octupole deformation $\beta_{33}$ superimposed
 on the oblate shape along  its symmetry axis. A result of this superposition
  is an oblate spheroid with a slightly triangular equator. The minimum
 corresponding to such a nuclear shape was previously reported for
 $^{308}$126 in \cite{JKS2010}. In contrast to the case of $^{308}$126,
 some of the oblate-$\beta_{33}$ minima in nuclei depicted in Fig. 6 lie
 significantly lower than
 the oblate minima obtained when assuming the axial symmetry.

  The landscapes around the g.s. minima in nuclei $^{296}$123 and $^{305}$124
 are shown in Fig. 7 in three different projections: $(\beta_{20},\beta_{22})$,
  $(\beta_{20},\beta_{30})$ and $(\beta_{20},\beta_{32})$. As previously,
 these maps are obtained by minimizing over the remaining eight deformation
 parameters. The oblate-$\beta_{33}$ minima are lower by 720 and 530 keV,
 respectively, than the axially symmetric oblate minima. As there is no
 barrier between both, the previously found axially symmetric minima were
 spurious.

 One has to mention that the depth of the oblate-$\beta_{33}$ minima
 diminishes with increasing pairing strengths which is especially relevant
 in odd-$A$ and odd-odd nuclei.
  As we have checked, at least some of these minima survive even after
 a 10\% increase in pairing strengths which corresponds to a considerable
 increase in rather weak g.s. pairing correlations of the original model.
  For example, such a change leads to the $\beta_{33}$ g.s. in $^{296}$123
  still lying by 450 keV lower than the axially symmetric minimum.

\section{Summary and conclusions}

 The results obtained for about 3000 heavy and superheavy nuclei by the
 microscopic-macroscopic model based on the deformed Woods-Saxon potential
 and the Yukawa-plus-exponential energy within the ten-dimensional space of deformations may be
 summarized as follows:

 - We could not find any deep minima of large tetrahedral deformation.
 The conditional minima, found under the restriction of zero quadrupole
 distortion, have mostly a large excitation and are not protected by
 any substantial barrier. The g.s. minima relatively
 soft with respect to the tetrahedral coordinate $\beta_{32}$ occur
  in Po isotopes with $N\approx 136$ and in a few very exotic (off $\beta$ -
 stability) systems.

 - The tetrahedral deformation $\beta_{32}$ appears in the g.s. minima when
  one combines it with $\beta_{30}$ and allows simultaneously for the
  quadrupole nonaxiality $\beta_{22}$. Then it turns out that
 in $\sim 40$ superheavy nuclei with $Z=119-126$, $N=173-188$, the ground
 states have a combined oblate and octupole deformation of the
 $\beta_{33}$ symmetry with respect to the axis of the oblate shape. The
  maximal g.s. lowering by this deformation, by 730 keV, occurs for the
 nucleus $^{296}$123.
  The effect, although reduced (to 450 keV in $^{296}123$), survives in the
 calculation with 10\% larger pairing strengths. This suggests
 some robustness of the prediction of oblate-$\beta_{33}$ ground states.

  Summarizing, one may thus say that our search for tetrahedral minima lead us
  instead to finding a combined oblate-plus-$\beta_{33}$ g.s. deformation
  in a restricted region of superheavy nuclei.

\section*{Acknowledgements}
 M. K. and J. S. were co-financed by the National Science Centre under Contract No. UMO-2013/08/M/ST2/00257
(LEA COPIGAL). One of the authors (P. J.) was co-financed by Ministry of Science and Higher Education: Iuventus Plus
grant nr IP2014 016073. This research was also supported by an allocation of advanced computing resources provided by
the \'Swierk Computing Centre (CI\'S) at the National Centre for Nuclear Research (NCBJ) (http://www.cis.gov.pl).

\end{document}